\documentclass[final,5p,times,twocolumn]{elsarticle}

\usepackage[utf8]{inputenc}
\usepackage{amssymb}
\usepackage{amsmath, amsthm, amssymb}

\usepackage[nodots]{numcompress}

\journal{Carbon}

\begin{document}

\begin{frontmatter}



\title{The microscopic origin of the hysteretic hydrogen adsorption in Double Wall Carbon Nanotubes bundles}


\author[CSIC]{Carlos Cabrillo\corref{cor1}}
\author[CSIC]{Ricardo Fern{\'a}ndez-Perea}



\author[MPC,DIPC,IKERBASQUE]{Felix Fern{\'a}ndez-Alonso}

\address[CSIC]{Instituto de Estructura de la Materia, (IEM), CSIC, Serrano 123, 28006 Madrid, Spain.}
\address[MPC]{Materials Physics Center, CSIC-UPV/EHU, Paseo Manuel de Lardizabal 5, 20018 Donostia - San Sebastian, Spain}
 \address[DIPC]{Donostia International Physics Center (DIPC), Paseo Manuel de Lardizabal 4, 20018 Donostia - San Sebastian, Spain}
 \address[IKERBASQUE]{IKERBASQUE, Basque Foundation for Science, Plaza Euskadi 5, 48009 Bilbao, Spain}

\cortext[cor1]{Corresponding author. Tel: +34 915616800, Email: ccabrilo@foton0.iem.csic.es (Carlos Cabrillo)}

\begin{abstract}
In a recent study we reported large hysteretic hydrogen adsorption in samples made out of Double Wall Carbon Nanotubes (DWCNT) at supercritical temperatures (50 K, 77 K and 150 K), a physical behavior of practical interest. Here, we center our efforts in elucidate its physical origin by means of neutron diffraction (ND) and computer simulations. After a detailed theoretical modeling of the ND signal we were able to quantitatively study how the nanostructure of the material reacts to the hydrogen loads. An expansion of the bundle structure with increasing hydrogen load was found. Taking advantage of the translational symmetry of a DWCNT bundle along the nanotube axis, a two dimensional numerical analogue of the system helped us to clarify the microscopic origin of the hysteresis. Two microscopic mechanisms are identified as responsible for the metastable hydrogen adsorption, one structurally static in character, namely, H$_{2}$-load dependent effective energy barriers between interstitials, the other related to the structural deformation, namely, the opening/closing of the channels between interstitials with increasing/decreasing H$_{2}$ loads owing to the concomitant bundle expansion/contraction. 
\end{abstract}

\end{frontmatter}



\section{Introduction}
Carbon nanotubes are paradigmatic of the nanoscience emergence during the last decades \cite{Hughes2024}. In particular, double wall carbon nanotubes (DWCNT) \cite{Endo2005, Kim2014} are nowadays object of notable efforts regarding its synthesis and functionalization \cite{IsidroGomez2026,Yang2026,Uceta2026,Hwang2025} given  its ample panoply of applications in areas, such as, batteries \cite{Zhao2024}, neuroscience \cite{Barrejon2022}, electrochemistry \cite{ThiThanh2023} or environment protection \cite{Wang2022}. Substantial hydrogen adsorption has also been reported by Miyamoto et al \cite{Miyamoto2006} (surpassing single wall counterparts \cite{Lyu2020}) and recently by ourselves \cite{Cabrillo2024}. Besides, we found a remarkable hysteretic behavior in the reported isothermal adsorption. 

Metastable adsorption due to capillary condensation is a well known phenomenon occurring at the mesoscale ($\sim$ 2 - 50 nm porous width) \cite{Rouquerol1999}. Its most characteristic feature is an hysteresis loop that sets in from below at a definite value of the pressure in units of the corresponding saturated vapor pressure,  $p/p_{0}$ ($\sim$ 0.4 in the case of N$_{2}$). There are, however, porous materials where the hysteresis loop does not close at low pressure and even  happens at supercritical temperatures. Such hysteresis cannot be understood in terms of condensation to the liquid phase. In the case of molecular hydrogen adsorption, the main interest is at supercritical temperatures (typically liquid nitrogen or room temperatures) since it comes from its potential to alleviate the high pressures needed to hydrogen storage in transport applications. Although not very common, hydrogen hysteretic adsorption has been previously reported in the context of Metal-Organic Frameworks (MOFs) as adsorbents \cite{Ferey2003,Zhao2004,Choi2008,Yang2009,Park2010,Yang2011,Zhong2011,Sircar2011,Dau2014}. It opens remarkable possibilities in energy applications since it can be used to lower the working pressure for a given uptake \cite{Zhao2004} or to facilitate thermal management (see Ref. \cite{Mason2015} for a case with methane). In general the supercritical hysteresis has been associated with flexible nanostructures that react to the molecular uptake in different ways \cite{Coudert2013}. But it has been associated also to molecular-size pore windows and restricted diffusion in rigid structures \cite{Kang2010,Sircar2011}.

In the case of nanostructured carbons, before our results, hydrogen hysteretic adsorption was found at room temperature and at low H$_{2}$ uptakes. The hysteresis was associated in this case with chemisorption in metallic impurities \cite{Barghi2014}. In counter opposition, our samples, made out of bundles of DWCNTs, display hysteretic hydrogen adsorption  at cryogenic temperatures ($\sim$ 3 wt\% at 1.3 MPa and 77 K) \cite{Cabrillo2024}. By means of elastic and inelastic neutron scattering we could characterize the structure of the material quantitatively (elastic scattering). Besides, at moderate H$_{2}$ loads both the structural information (elastic scattering) as well as the microscopic diffusive dynamics (quasielastic scattering) identified the interstitials between bundles as the main adsorption sites for the molecular hydrogen. Likewise, the rotational dynamics of the adsorbed H$_{2}$ molecules (inelastic scattering) unambiguously discarded the possibility of any significant chemisorption at metallic impurities.

Once ascertained the physisorptive character of the adsorption, the question arises about the microscopic origin of the hysteresis and how it fits within the framework of the previously reported observations. With such a goal in mind we performed a series of measurements of the system and how it reacts under hydrogen adsorption by making recourse to a neutron diffraction instrument with a wide range of scattering vector magnitude. A detailed model of the observed signal has allowed us to scrutinize how the nanostructured material reacts to the hydrogen load. In order to gain the physical insight sought for, the measurements have been complemented with numerical simulations (mainly molecular dynamics) of a two dimensional analogue of the real material. Such a simplification, albeit drastic, allows for reasonable computational loads while, owing to the translational symmetry of a DWCNT bundle, retaining key aspects fo the adsorption dynamics of the real system  avoiding the prohibitively high computational loads of atomistic simulation of the real material. Altogether has allowed us to unveil the microscopic mechanisms responsible of the observed hysteretic behavior, a behavior that, as mentioned above, its unique in the realm of carbon nanostructures materials.  

The present report is structured as follows. In section \ref{MandM}, both the experimental and the numerical simulation details are given. Here is also developed a real space theoretical model of the neutron diffraction (ND) measurements. In section \ref{RandD} the model is first applied to the pristine material, the results showing consistency with our previous studies. The model happens to be accurate enough to permit proper fits of the ND measurements under increasing loads of molecular hydrogen at 50 K and 77 K revealing a structural expansion of DWCNT bundles induced by the hydrogen adsorption. The simulations of the two dimensional analogue are also presented and the various results are discussed. Finally, the report is ended with a summary and a compilation of the main conclusions in section \ref{SandC}.

\section{Experimental and theoretical}\label{MandM}

\subsection{Materials}

All the DWCNT samples here studied were taken as such from a 1 g of material made by Materials \& Electrochemical Research (MER) Corporation labeled as Catalytic Double-wall C-nanotube, purified, 5 wt\% Fe. The original material takes the form a of rather thick, somewhat fluffy, black fabric that resembles the so called Buckypaper also made out of DWCNTs \cite{Endo2005}. Details about the sample characteristics can be found in \cite{Cabrillo2024}.

\subsection{Neutron scattering}\label{NS}
Neutron diffraction experiments were performed on the NIMROD instrument, a total scattering diffractometer at the ISIS facility, Rutherford Appleton Laboratory, UK \cite{NIMROD}. NIMROD is a unique time-of-flight (TOF) neutron instrument~\cite{Bowron2010}. By virtue of its design using pulsed spallation neutrons, it spans a very wide range of momentum transfers ($0.02 \leq Q \leq 100$~$\AA^{-1}$) with an exceptionally high counting rate. Such a wide $Q$ range is ideally suited to explore mesoscopic length scales involved in molecular intercalation processes ($\sim$ 1 nm).

The sample holder was a flat bespoke cell of linear internal dimensions 56 mm height, 35 mm width, and 3 mm depth with 3.5 mm wall thickness capable of withstanding up to 80 bar of pressure. Made of a TiZr null-scattering alloy, the holder scatters neutrons only incoherently so that do not add Bragg peaks to the measured signal. The cell was connected to a centerstick adapted for use with a 4-K closed-cycle refrigerator (CCR) and to a gas-handling manifold (standard facility equipment) by use of stainless-steel tubing (4 mm inner diameter). The pressure within the system was monitored with a MKS 627B manometer (10000 Torr range, accuracy of 0.12\% of the reading) as well as with pressure transducers of up to 60 bar and 200 bar. Calibrated thermocouples (RhFe) and heaters were placed at the top and bottom of the sample cell. This experimental setup allowed for fine temperature control ($<$0.1~K) over the range 4-400~K.

Prior to the Neutron Diffraction (ND) experiments, the pristine DWCNT material was dry in a vacuum oven five days, the first three days at 200 C$^{\circ}$, the fourth 100 C$^{\circ}$ and a final day at 60 C$^{\circ}$. A sachet made out of aluminium foil 180 $\mu$m width was prepared and filled with 339.35 mg of the DWCNT material and quickly introduced in the sample holder and sealed. Before introducing the sachet into the holder, ancillary ND measurements of the empty cell (needed for the estimate background) and a vanadium sample (needed for normalization) were made.

We first made a made a ND measurement of the pristine material at room temperature. After equilibrating the system at 77 K, we performed a series of measurements following an ascending/descending ramp of H$_{2}$ pressures up to a nominal maximum of 50 bar. The same protocol was repeated at 50 K. All the data treatment and corrections (except from inelasticity) such as conversion from TOF to energy, detector calibration, shelfscattering, multiple scattering, etc, were made with the Gudrun software package \cite{gudrunweb}. In order to carry out the corrections the composition of the material is used. On the other hand, Gudrun also yields by angular integration the expected total scattering cross-section per atom which must coincides (within a 10 \%) with that calculated from the known composition. By selfconsistent addition of H$_{2}$ molecules to the theoretical composition, the excess of H$_{2}$ irradiated by the beam during the hydrogen loaded ND runs can be estimated \cite{Cabrillo2016,Cabrillo2019} and applied to the correction in a second iteration of Gudrun. There is, however, a major correction, that is, the inelasticity of the scattering from the hydrogens present in the sample, not implemented in this procedure. Gudrun incorporates an advanced procedure for such a task \cite{Soper2009,gudrunweb} but it does not apply well to samples that scatter strongly in the small angle region as it is in our case. Instead, following gudrun alternative option, we have used an heuristic approach consisting in adding decaying exponential fittings to the data. In practice, in the $Q$ region of interest, just one exponential decay plus a constant background are enough, added to the modeling fits outside gudrun.   

Given the tight schedule of beam time in a neutron scattering facility we could not made all the necessary ancillary measurements (some of them requiring quite long equilibration times) to perform H$_{2}$ adsorption measurements concomitant with the ND measurements.

With respect to the resolution of our measurements a TOF instrument such as NIMROD present an angular dependence of the relative resolution in momentum transfer, $\Delta Q/Q$, which is relatively constant at large angles but increase quickly at low angles (below ten degrees in the case of NIMROD \cite{Bowron2010}). In our case, our interest is centered around a peak at $\sim 0.3$ \AA$^{-1}$ and a shoulder around $\sim 0.6$ \AA$^{-1}$. In order to increase the statistics sufficiently, Gudrun mixes data coming from detectors at different angles including those below ten degrees making the estimation of the resolution difficult. However, we could resolve the shoulder using only detectors from angles of minimal relative resolution ($\sim 4$ \%). From the fitted feature assuming a $\Delta Q/Q=0.04$ an increase up to $\Delta Q/Q=0.12$ was needed with a Gaussian resolution profile in order to reproduce the shoulder when all detectors are included. Besides, we eventually modeled the peak and the shoulder leaving the resolution as fitting a parameter again obtaining $\Delta Q/Q=0.12$. Since what is constant is the relative resolution, the effect of the resolution is accounted for by a running integral with a ``Gaussian" of linearly variating dispersion $\sigma = 0.12 \, Q$ instead of a standard convolution.

\subsection{A real space modeling of the neutron diffraction signal}\label{ElasNSmodel}

In our previous work \cite{Cabrillo2024}, following previous reports, we relied on the time honored reciprocal space approach to develop a model of the observed (elastic) neutron diffraction. Although we got satisfactory fits, this approach presents some caveats regarding some aspects of the system physical description. In particular, it assumes a crystalline order of the 2D triangular array of nanotubes within a bundle, i.e., a perfect triangular lattice on average for the locations of the tube centers. Displacements (thermal for instance) around such crystalline lattice give raise to a Debye-Waller factor that decreases the intensity of the diffraction features as $Q$ increases but do not produce a broadening of the diffraction features. A broadening is then ascribed only to a finite size of the crystals (crystallites). This kind of order assumes that there is no dispersion on the first neighbor distance during the `building" of the crystal. This is not our case: the nanotubes present a dispersion in the diameters giving rise to a dispersion in the pair distances that increases (by accumulation) with the distance in such a way that there is no perfect lattice even on average. This so called paracrystalline order induces broadening of the diffraction peaks even for infinite arrays (see chapter 9 in \cite{Guinier1994}).

Ideally, in a the total scattering TOF diffractometer, the scattering per atom coming out of a collection of $N$ atoms of the same element with null incoherent neutron cross section (as is the case of carbon) is proportional to,
\begin{equation}
\label{NS}
S(\mathbf{Q}) = 1 +  \frac{1}{N}\sum_{i \neq j}  \left \langle \exp{\left[i (\mathbf{t}_{i}-\mathbf{t}_{j}) \cdot
\mathbf{Q} \right]}  \right \rangle \, ,
\end{equation}
\noindent
where $ \mathbf{t}_{i}$ is the position vector of the ith-nucleus, $\mathbf{Q}$ is the scattering vector (elastic momentum transfer, i.e., at zero energy transfer) and $\langle \rangle$ represents thermal average. In practice, as described in the previous section, a series calibrations and corrections onto the measured TOF differential cross section are needed to approach Eq.~(\ref{NS}). As we will see later these corrections will be of particular relevance in our case. Implicitly, the measurements implement also sample averages over the various parameter distributions defining the material such as diameters of the nanotubes, orientation, etc, so that $\langle \rangle$ can be reinterpreted not only as a thermal average but as ensamble averages over those parameters of an underlying model.  

Let us now decompose $\mathbf{t}_{i}$ in a component corresponding to the center of the nanotubes, $\mathbf{R}_{\alpha}$, and intratube position vector, $\mathbf{r}_{i(\alpha)}$, so that $\mathbf{t}_{i} = \mathbf{R}_{\alpha} + \mathbf{r}_{i(\alpha)}$,  being all the $\mathbf{r}_{i(\alpha)}$ in the same plane perpendicular to the tube axis. Setting apart the intratube terms, Eq.~(\ref{NS}) reads,
\begin{multline}
\label{NSintra}
S(\mathbf{Q}) = 1 +  \frac{1}{N}  \bigg \langle \sum_{\alpha}\sum_{i \neq j}  \exp{\left[i (\mathbf{r}_{i(\alpha)}-\mathbf{r}_{j(\alpha)}) \cdot
\mathbf{Q} \right]} \,+ \\
\sum_{\alpha \neq \beta} \exp{\left[i (\mathbf{R}_{\alpha}-\mathbf{R}_{\beta}) \cdot
\mathbf{Q} \right]}  \sum_{i(\alpha), j(\beta)} \exp{\left[i (\mathbf{r}_{i(\alpha)}-\mathbf{r}_{j(\beta)}) \cdot
\mathbf{Q} \right]} \bigg \rangle \, .
\end{multline}
\noindent
The intratube part in this expression is proportional to a sample average over the tubes. Likewise, the intertube part can be understood as a sample average over bundles, not made explicit in Eq.~(\ref{NSintra}) just for not to burden the notation. As usual, we will assume that the various averages are uncorrelated, a first approximation that greatly simplify the analytical modeling. Taking also into the account that the thermal average gives rise to a Debye-Waller factor Eq.~(\ref{NSintra}), can be written as,
\begin{multline}
\label{NSintra2}
S(\mathbf{Q}) = 1 +  \frac{e^{-2W(\mathbf{Q})}}{N} \Bigg\{N_{t}\bigg \langle \sum_{i \neq j}  \exp{\left[i (\mathbf{r}_{i}-\mathbf{r}_{j}) \cdot
\mathbf{Q} \right]} \bigg \rangle \,+ \\
N_{b}\bigg \langle \sum_{\alpha \neq \beta} \exp{\left[i (\mathbf{R}_{\alpha}-\mathbf{R}_{\beta}) \cdot
\mathbf{Q} \right]} \bigg \rangle \bigg \langle \sum_{\alpha \neq \beta } \exp{\left[i (\mathbf{r}_{i(\alpha)}-\mathbf{r}_{j(\beta)}) \cdot
\mathbf{Q} \right]} \bigg \rangle \Bigg\} \, , 
\end{multline}
\noindent
where the argument of the Debye-Waller factor, $2W(\mathbf{Q})$, will be determined later, $N_{t}$ is the number of nantotubes and $N_{b}$ the number of bundles. Defining now a structure factor and a kind of precursor of a form factor as,
\begin{eqnarray}
S_{2D}(\mathbf{Q}) & = &  \frac{1}{\overline{N}_{tb}}\sum_{\alpha, \,\beta} \bigg \langle \exp{\left[i (\mathbf{R}_{\alpha}-\mathbf{R}_{\beta}) \cdot 
\mathbf{Q} \right]} \bigg \rangle \, , \\
F(\mathbf{Q}) & = & \frac{1}{\sqrt{\overline{N}_{c}}}\sum_{i}  \exp{\left[i \mathbf{r}_{i} \cdot
\mathbf{Q} \right]} \, ,
\end{eqnarray}
Eq.~(\ref{NSintra2}) reduces to,
\begin{multline}
\label{NSintra3}
S(\mathbf{Q}) = 1 - e^{-2W(\mathbf{Q})} + e^{-2W(\mathbf{Q})}\bigg \{ \langle | F(\mathbf{Q})  |^{2} \rangle-| \langle F(\mathbf{Q}) \rangle  |^{2} \,+ \\ 
 | \langle F(\mathbf{Q}) \rangle  |^{2} \, S_{2D}(\mathbf{Q}) \bigg\} \, ,
\end{multline}
\noindent
where $\overline{N}_{tb}$ is the mean number de nanotubes in a bundle and $\overline{N}_{c}$ is the mean number of carbon atoms in a nanotube. The first part, i.e., $1 - e^{-2W(\mathbf{Q})}$ corresponds to the diffuse scattering owing to the thermal fluctuations. $S_{2D}(\mathbf{Q})$ is the structure factor corresponding to the two-dimensional triangular lattice of nanotube positions in a bundle. Notice that, under the assumption $ \langle | F(\mathbf{Q})  |^{2} \rangle \approx | \langle F(\mathbf{Q}) \rangle  |^{2} $ the part between curly brackets in Eq.~(\ref{NSintra3}) reduces to the standard approximation of a form factor by a structure factor. Here, instead, we have a form factor, $ | \langle F(\mathbf{Q}) \rangle  |^{2}$, and a background given by the variance of $F(\mathbf{Q})$.

In the $Q$ range of interest, the nanotube structure is well approximated by a continuos distribution with areal density $\rho_{A} =  4/(3 \sqrt{3} \, d_{cc}^{\,2})$, where $d_{cc}$ is the distance of the C$-$C bond
in graphene (1.42 \AA). The sum in $F(\mathbf{Q})$ can then be substituted by an integral. Taking the limit of infinite nanotube length and performing an isotropic orientational average, the square of $F(\mathbf{Q})$ goes as $\sim R^2 J_{0}(Q R)^2/Q$ being $R$ the radius and  $J_{0}$ the first kind Bessel function of  order zero. In the case of a DWCNT it will a sum of such a term plus the corresponding to a radius $ R + h$ being $h$ the distance between the inner and the outer tubes. However, there are bundles mixing both DWCNTs and SWCNTs. In order to account for such a possibility we define and effective form factor corresponding to a combination of both possibilities with a proportion $x_{1w} \in [0,1]$, with $x_{1w} =1$ for purely SWCNTs and $x_{1w} =0$ for DWCNTs. All together 
yields the following expressions,
\begin{multline}
\label{FF1}
| \langle F(Q) \rangle  |^{2} =\frac{2 \, \pi^{2}\rho_{A}}{(2  \overline{R} +h) Q} \\
\Big [ \langle R J_{0}(Q R) \rangle + (1-x_{1w})\langle (R + h)J_{0} \big (Q (R + h) \big ) \rangle \Big ]^{2} \, ,
\end{multline}
\noindent
and,
\begin{multline}
\label{FF2}
\langle | F(Q)  |^{2} \rangle =\frac{2 \, \pi^{2}\rho_{A}}{(2 \overline{R} +h) Q} \\
\left \langle \Big [  R J_{0}(Q R) + (1-x_{1w})(R+h)J_{0} \big (Q (R + h ) \big ) \Big ]^{2} \right \rangle \, ,
\end{multline}
\noindent
where $ \overline{R}$ is the mean nanotube radius and now $\langle \rangle$ represents the ensamble average only over the radii distribution (we consider $h$ dispersionless). By virtue of the infinite tube length limit, $\mathbf{Q}$ is now restricted to the perpendicular plane to the nanotube axis and $Q$ is its modulus. 

In regarding $S_{2D}(\mathbf{Q})$, decomposing $\mathbf{R_{\alpha}}$ in the mean value plus a fluctuation $\mathbf{R_{\alpha}}=\overline{\mathbf{R}}_{\alpha}+\delta \mathbf{R}_{\alpha}$ and assuming an isotropic Gaussian distribution for $\delta \mathbf{R}_{\alpha}$ we have,
\begin{equation}\label{Sq2D}
S_{2D}(\mathbf{Q})  =  1+ \frac{e^{-\sigma_{\alpha \beta}^{2} \,Q^{2}}}{\overline{N}_{tb}}\sum_{\alpha \neq \beta} \bigg \langle \exp{\left[i (\overline{\mathbf{R}}_{\alpha}-\overline{\mathbf{R}}_{\beta}) \cdot 
\mathbf{Q} \right]} \bigg \rangle \, ,
\end{equation}
\noindent
where $\sigma_{\alpha \beta}$ is the dispersion in $|\overline{\mathbf{R}}_{\alpha}-\overline{\mathbf{R}}_{\beta}|$. After orientational average, the 2D structure factor reads,
\begin{equation}\label{Sq2Dor}
S_{2D}(Q)  =  1+ \frac{e^{-\sigma_{\alpha \beta}^{2} \,Q^{2}}}{\overline{N}_{tb}}\sum_{\alpha \neq \beta} J_{0}\left  ( |\overline{\mathbf{R}}_{\alpha}-\overline{\mathbf{R}}_{\beta}| \, Q \right )  \, .
\end{equation}
\noindent
Owing to the paracrystalline nature of the 2D lattice, $\sigma_{\alpha \beta}$ cannot be interpreted as in a Debye-Waller factor, i.e., a characteristic constant of the crystal. Instead, the dispersion depends on the pair distance. To a good approximation it can be considered proportional to the squared root of the distance (exact for 1D chains) \cite{Cabrillo2011} so that it can be expressed in terms of the dispersión of the first neighbor distance.  In our case it amounts to,
\begin{equation}\label{paradis}
\sigma_{\alpha \beta}^{2} =  (2\sigma_{R}^{2}+\sigma_{g}^2) \, \frac{ |\overline{\mathbf{R}}_{\alpha}-\overline{\mathbf{R}}_{\beta}|}{2 \overline{R} + \overline{g}} \,  ,
\end{equation}
\noindent
where $\sigma_{R}$ is the dispersion in the radii distribution, $\overline{g}$ is the mean value of the gap between carbon nanotubes and $\sigma_{g}$ its dispersion. In variance with a crystal, the dispersion now grows with $ |\overline{\mathbf{R}}_{\alpha}-\overline{\mathbf{R}}_{\beta}|^{1/2}$ so that eventually it becomes larger than the separation between particles (first neighbor distance, i.e., $2 \overline{R} + \overline{g}$).  As a consequence, long-range order is lost and the Bragg peaks broadens.

To perform the sum in $S_{2D}(Q)$ we rely in the discrete pair distribution function, that is, a lookup table with entries of the form $g_{n} = \{d_{n}, m_{n}\}$ where the $d_{n}$'s are the $|\overline{\mathbf{R}}_{\alpha}-\overline{\mathbf{R}}_{\beta}|$'s distances arranged in increasing order and $m_{n}$'s are the corresponding multiplicity. The distances are expressed in terms of  the lattice constant $d = 2(\overline{R} +(1-x_{fn}) h + \overline{g}\,)$, i.e., $d_{n} = d t_{n}$, where, following \cite{Cabrillo2021}, $x_{fn}$ is not identified with $x_{1w}$ but left as a fitting parameter and the $t_{n}$'s are the distances of a finite triangular lattice of unit lattice constant. The $t_{n}$'s lookup table can be calculated very efficiently for very large sizes. Once normalized to the maximum pair distance, the ratio between the multiplicities of a finite crystal and those in the thermodynamic limit (for practical purposes a sufficiently large reference crystal) follows a universal curve independent of the size \cite{Cabrillo2011}. Then, from the lookup table of the reference crystal, the pair distribution of any finite size crystal can be obtained through multiplication by the master curve. We used such an approach for a first estimation of the size of the 2D crystallites. Not surprisingly, the estimated sizes resulted rather small. Adding the 2D character of the crystals, an exact computation of the $g_{n}$'s are computationally cheap so that a list of the exact $g_{n}$'s corresponding to crystallites of increasing sizes were used in the final calculations. 

After all corrections and calibrations gudrun yields the interference part of the diffraction signal which amounts to $S(Q) - 1$ multiplied by an instrumental scale factor. Besides there is a Porod contribution, i.e., a tail of the form $1/Q^{4}$, to the small angle scattering coming from the meso and macro scale estructures present in the sample \cite{Porod1951,Guinier1994Chap10}. Altogether yields for our final model,
\begin{multline}
\label{IQ}
I(Q) = A_{inst} \, e^{-2W(Q)}\bigg \{ \langle | F(Q)  |^{2} \rangle-| \langle F(Q) \rangle  |^{2} -1  \,+ \\ 
 | \langle F(Q) \rangle  |^{2} \, S_{2D}(Q) \bigg\}\otimes R(Q) +\frac{A_{P}}{Q^{4}}+A_{in} e^{- \gamma \, Q} +bck \, ,
\end{multline}
\noindent
where $A_{inst}$, $A_{P}$, $A_{in}$, are the instrumental, Porod law and inelastic correction amplitudes respectively. The inelastic correction is further determined by $\gamma$ the decay and constant background, $bck$, as discussed above. $R(Q)$ is the resolution function, described in the preceeding subsection. As observed there, the symbol $\otimes$ do not denote here the standard convolution since the dispersion of $R(Q)$ is not constant but linear in $Q$. To simplify calculations the Porod and inelasticity terms has been left outside the resolution action since the effect is negligible on such terms. As in \cite{Cabrillo2024} the average over radii, $\langle \rangle$, has been implemented by drawing radii from a Gamma distribution of mean, $\overline{R}$ and polydispersity $\sigma_{p} = \sqrt{\langle (R -\overline{R})^2\rangle}/\overline{R}$, our last fitting parameter. 

\subsection{Fitting procedure}
To compare the model against measurements the NonlinearModelFit function of the Mathematica package which explore the parameter space in search of the best fit, yielding vast information about errors and quality of the fits  has been used. Some of the parameters, however, have been scanned manually in search of the best fit owing to the discrete nature of the parameter (number of nanotubes in a bundle) or failure to converge in a reasonable computing time. The Bayesian Criterium parameter (BIC) and the Coefficient of determination, R$^{2}$, has been used to ascertain the goodness of the fits. 

\subsection{Numerical simulations}\label{CS}
Complementary to the ND measurements we have performed to types of Molecular Dynamics simulations (MD), a standard atomistic simulation of the H$_2$-bundle system and a two-dimensional analogue. All the MD simulations have been carried out using the Large-scale Atomic/Molecular Massively Parallel Simulator (LAMMPS) code \cite{Thompson2022}. 

\subsubsection{Atomistic simulations}\label{CSa}
Within a simulation box of $222.54 \times 251.80 \times 480.5$ Å$^3$ a bundle of nineteen (10,10)@(15,15) DWCNTs were located in a triangular array with hexagonal symmetry aligned along the z axis (see Fig.~\ref{2DScheme} for a schematic section normal to the z axis). This simulation box accommodates 195 (10,10)@(15,15) primitive cells thus allowing substantially large wavelength vibrations of the nanotubes once periodic conditions are applied. Likewise, in the plane xy the box is large enough to avoid self interaction of the bundle under periodic conditions. The radius of the chosen DWCNT is a compromise between those corresponding to our own adsorption measurements \cite{Cabrillo2024} and previous studies in DWCNT bundles showing similar hydrogen adsorption \cite{Miyamoto2006}. Hydrogen was loaded in the simulation box into the space outside the bundle. A battery of simulations with hydrogen loads equivalent to pressures in the empty simulation box between 0.1 MPa to 10 MPa, ranging form 5 K to 800 K in temperature and with simulation times between 2 nanoseconds and 150 nanoseconds were performed. All the interaction between the atomic species were implemented with the Adaptive Intermolecular Reactive Empirical Bond Order potential (AIREBO) \cite{Stuart2000}. 

\subsubsection{Two-dimensional simulations}\label{CS2D}
In this simplified model the 3D space is reduced to a plane perpendicular to the nanotubes where the nanotubes and the H$_{2}$ molecules are represented by disks. Again nineteen disks in a triangular lattice with an hexagonal arrangement represent a bundle of DWCNT's (inset Fig.~\ref{2DScheme}) this time located within a simulation box of $200 \times 200$ Å$^2$. To choose the paramaters of the this 2D analogue we matched as much as possible the atomistic 3D simulation. Thus, the radii of the nanotube disks was 10.21 Å and the mass 1800 Da corresponding to the radius and mass of a primitive unit cell of a (10,10)@(15,15) DWCNT (the thinnest slice of such a double nanotube). From the 3D simulations we learnt that two neighbors nanotubes confronts approximately three carbon hexagons of their perimeters which, assuming an AB graphitic disposition, amounts to nine first neighbors carbons confronted in each carbon in a primitive cell. Thus, we assumed an harmonic potential interaction between the nanotube disks with a strength constant of k = 88.038 meV Å$^2$ (nine times the strength of a C-C interaction between two graphitic layers) and an equilibrium distance of 23.82 Å giving rise to a gap between disks of 3.4 Å. To represent the hydrogen molecules disks of 2 Da of mass interacting between them by means of a Lennard-Jones potential with $\epsilon = 2.126$ meV as obtained by fitting to the Silvera-Goldman potential \cite{Silvera1978}. The interaction with the nanotube disks is also implemented through a Lennard-Jones with $\epsilon = 40$ meV as a round number derived from published results \cite{Mattera1980,Ghio1980,Ye1999}. The corresponding sigmas of the Lennard-Jones were set equal, and the common $\sigma$ used as a parameter characterizing the size of the molecule.

\begin{figure}
\centering
\includegraphics[width=0.48 \textwidth]{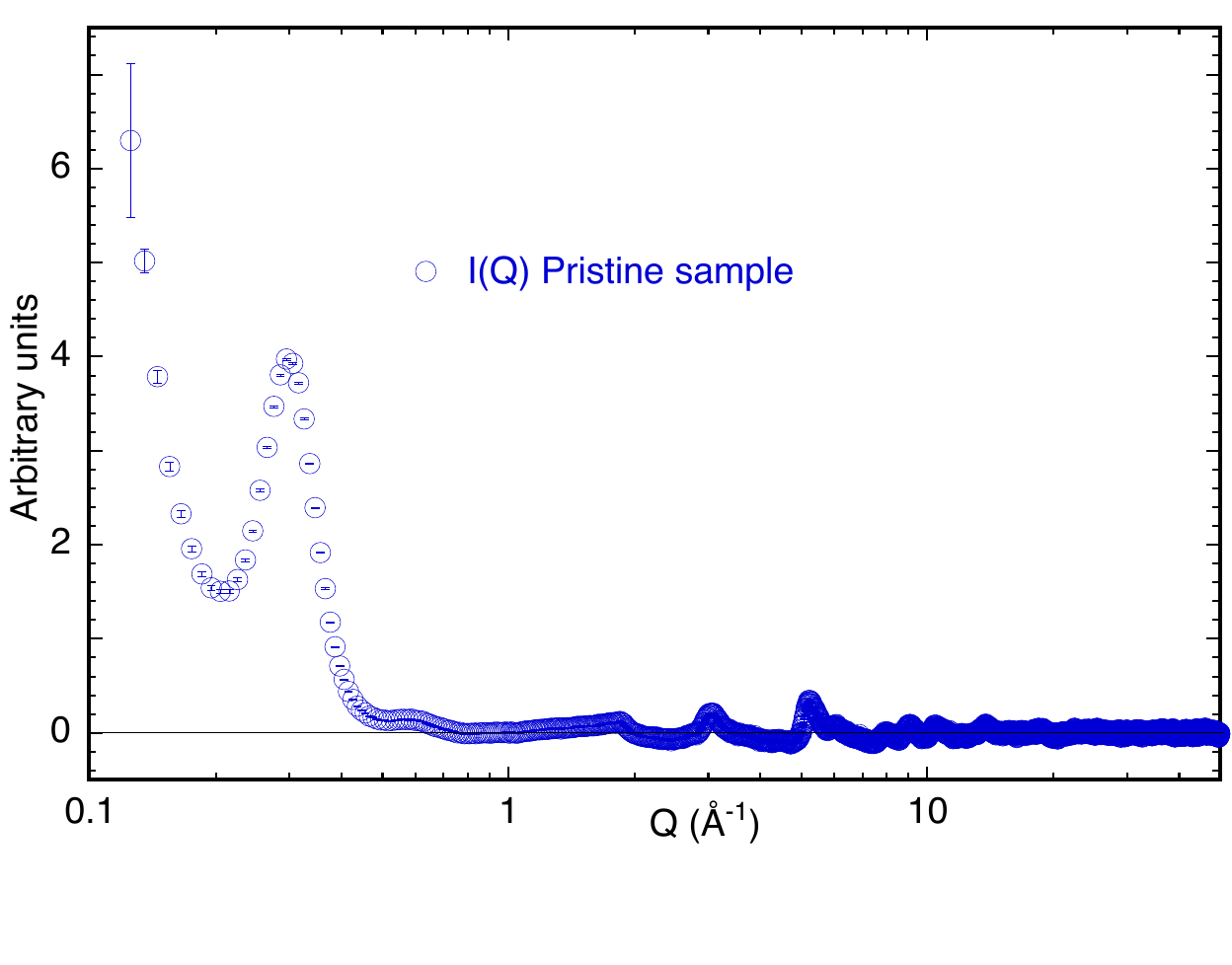}
\caption{\label{NDempty} Neutron diffraction of the pristine sample. Notice the logarithmic scale in the Q axis.}
\end{figure}
\noindent

With this system we simulated adsorption isotherms as a function of $\sigma$ at 50 K, 77 K and 150 K. In particular, for each adsorption step, a load of hydrogen disks was placed in outermost part of the square simulation box.  A MD simulation were run for as long as needed to stabilise the density of the ``hydrogen" disks in the outermost part of the simulation box and a equilibrium density was obtained from the final ns of the simulation. Given the small size of the system, in order to reduce equilibration time, independent thermostats were used for the ``molecular" disks and the ``nanotube" counterparts. Following the same procedure as in the real experiments, the difference between the number of disks present in the simulation box and the amount of disks that should be present in the free area of the box assuming the calculated density were uniform along the whole free area gives the adsorbed disks. The free area was obtained by subtracting the area occupied by the nineteen disk that conform the bundle augmented to take into account the effects of core particle repulsion. Thus, the radius at the minimum of the the interaction between helium and carbon (Lorentz-Berthelot mixing of conventional Lennard-Jones C-C and He-He interactions) was added to the disk radii amounting to an augmented radius of 13.5 Å. Helium is the gas for the standard pychnometry used to measure the free volume in the real experiments. Some simulations have been done with the nanotube disks fixed in their equilibrium positions. In such cases, the overlap between the augmented disks has been taken into the account for a proper estimation of the free area. To follow the experimental procedure the numerical isotherms are given in pressure units corresponding to the temperature by the areal disk density (gas constant taken as 1). 

In some cases, in order to avoid metastability and to reach the true thermodynamic equilibrium, we have relied in a metropolis Montecarlo approach moving the H$_{2}$ disks by large jumps of 13 Å, enough to overcome the gaps between nanotubes. Besides, every 3000 jumps, 20 ps of MD were run.

\section{Results and discussion}\label{RandD}
\subsection{Neutron diffraction measurements}
The overall ND measurement of the pristine sample is depicted in Fig.~\ref{NDempty}. The most conspicuous feature is the big peak at $Q \sim 0.3$ Å$^{-1}$ which is characteristic of the triangular arrangement of the nanotubes in bundles. It is the object of our main interests and its prominence speaks of a good quality sample. Likewise the absence of a prominent peak at around 1.7 Å$^{-1}$ but a weak feature instead, indicates low multiwall nanotubes and/or turbostratic impurities levels. We will center in the $Q$-range between 0.1 and 2 Å$^{-1}$ thus including also the shoulder immediately after the main peak since it carries out also information about the bundles. Our main concern is to elucidate how the adsorption of hydrogen alters the bundle structure, in particular, how it expands if so. Consequently, the most relevan parameter of the model is the gap between nanotubes within a bundle, $g$. Reassuringly the values of the parameters already explored in our previous study \cite{Cabrillo2024} worked perfectly. There are however parameters not present in our previous model. To begin with, the Debye-Waller factor was left to 1, that is, $2W(Q) = 0$. Here, we have set  $2W(Q) = \frac{1}{3} u^{2} Q^{2}$ with $u$ the root squared displacement with $u = 0.1$ Å as estimated from our atomistic simulations of a DWCNT bundle at 77 K. This value agrees well with the experimental values for graphene \cite{Shevitski2013}. Once fixed $2W(Q)$, we could perform stable, reasonably quick automatic fittings with as many as seven free parameters, namely, $A_{inst}$, $A_P$, $A_{in}$, $\gamma$, $bck$, $\sigma_{p}$, and $g$, the main object of our interest. The background, $bck$, was not needed in the pristine sample but necessary once H$_{2}$ was loaded. The model depends also on the size and the shape (weakly) of the bundle. Within this last respect lacking any simple parametrization of the shape, we focused on the most isotropic bundles in the normal section, that is, those built from a central nanotube up to a given radius. A table of fits with increasing bundle radii was computed and the goodness of the fits compared with the Bayesian Information Criterium (BIC) parameter. 

\begin{figure}[t]
\centering
\includegraphics[width=0.48 \textwidth]{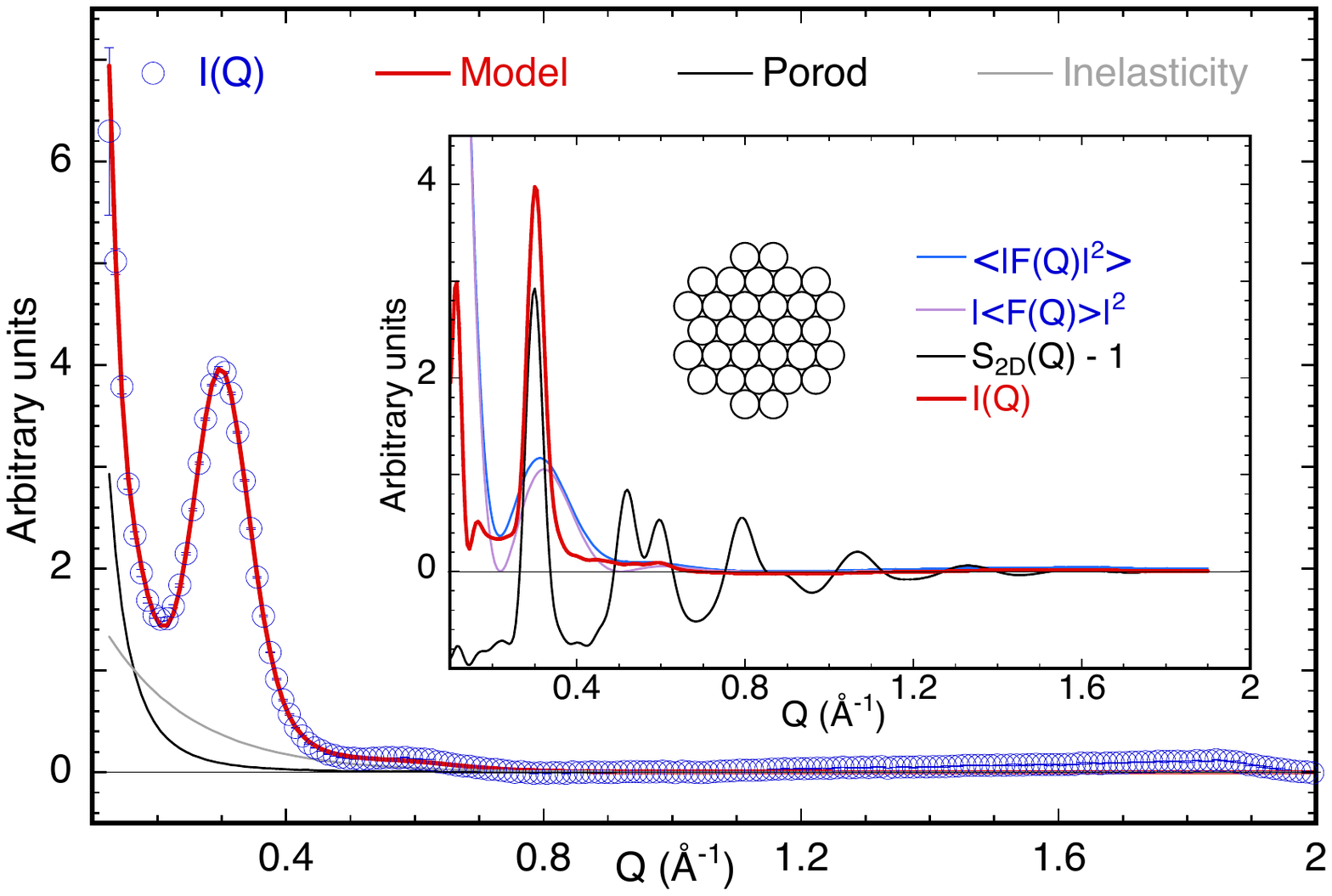}
\caption{\label{NDemptyModel} Neutron diffraction of the pristine sample along with our fitted model, the Porod law contribution and the hydrogen inelasticity correction. The inset shows the model and its main components after removing the experimental resolution as well as a normal section of the structure of the bundle.}
\end{figure}
\noindent

The results are summarized in Fig.~\ref{NDemptyModel} and Table \ref{DModTable}. The first thing to notice is that, in spite of our efforts, during the load of the sample in its holder a small proportion of water was adsorbed significant enough to need a hydrogen inelasticity correction. The fit capture very well the peak and its associated shoulder as well as the small angle scattering. Notice that our model gives a small angle signal coming from the bundle (as opposed to the reciprocal space method used in \cite{Cabrillo2024}) but, as expected, a further contribution is needed. Reassuringly, the parameters agreement with the reciprocal space method is also excellent. The only parameter that has been slightly refined is $\sigma_{p}$ while $g$ has been optimized using the NonlinearModelFit procedure, both comparing well with the previous results, namely, $\sigma_{p} = 0.2$ and $g = 3.2$ (see Table \ref{DModTable}). There was no necessity of further refinement of the rest of manually optimized parameters. There is, however, a clear qualitative difference, between the paracristalline real space approach followed here and the reciprocal space counterpart, namely, a substantially more ``crystalline" 2D structure factor $S_{2D}(Q)$ (see inset Fig.~\ref{NDemptyModel}). 

The size of the most representative bundle (see inset in Fig.~\ref{NDemptyModel}) corresponds to a radius of 2.7 nanotube diameters which amounts to maximum pair distance of 5.29 Å. Good fits are also obtained for the previous size of 2 diameters (hexagonal symmetry bundle). For smaller sizes the BIC parameter quickly increases. Good fits happens again for larger bundles, albeit increasing accordingly the $\sigma_{p}$ value. At any rate, the parameters here determined agree well with previously reported studies. More specifically, $g = 3.2$ Å have been reported in \cite{Abe2003} (here 3.1 Å) and $h = 3.6$ Å in \cite{Abe2003,Hwang2025}. With respect to the polydispersity $\sigma_{p} \sim 0.2$ is roughly half of that deduced from data in \cite{Hwang2025} what make sense since the DWNTs there, do not form bundles. A large polydispersity hinders bundle formation. 

\begin{table}[t]
\begin{center}
\begin{tabular}{cccccccc}
\hline
$ \langle R \rangle $ & $\sigma_{p}$  & $n_{R}$ & $h$ & $g$ & $x_{1w}$ & $x_{fn}$ \\ 
 \hline \noalign{\smallskip}
9.65 Å & 0.17 & 10$^{4}$ & 3.6 Å & $3.07 \pm 0.02$ Å & 0.4 & 0.75 \\
 \hline
\end{tabular}
\caption{Most relevant ND model parameters used in Figure \ref{NDemptyModel}. $ \left <R \right >$ is the mean inner radius, $\sigma_{p} $ the polydispersity of the inner radius gamma distribution, $n_{R}$ the number of sampled radii, $h$ is the spacing between the walls in DWCNTs, $g$ is the gap between tubes in a bundle, $x_{1w}$ is the proportion of SWCNTs in the effective form factor, $x_{fn}$ is the proportion of SWCNTs in the effective 2D triangular lattice. The values without errors has been scanned manually.}
\label{DModTable}
\end{center}
\end{table}

\begin{figure}[t]
\centering
\includegraphics[width=0.48 \textwidth]{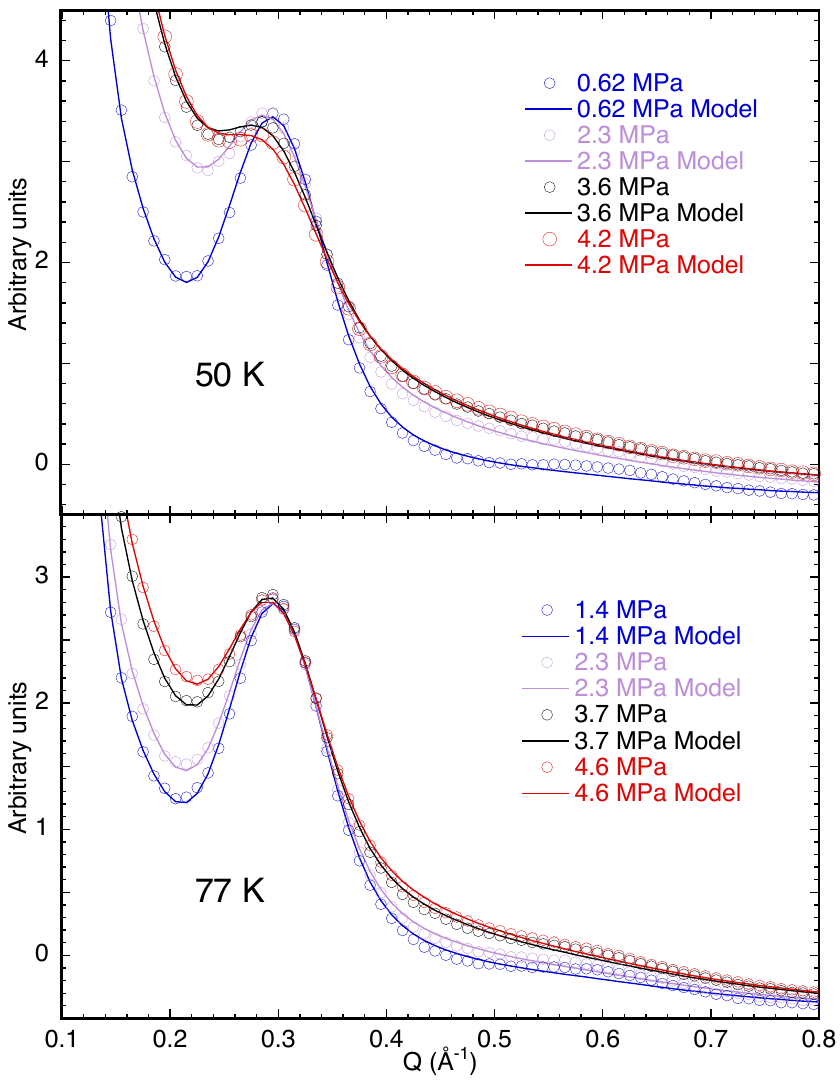}
\caption{\label{NDH2loads} Selected neutron diffraction measurements at four increasing H$_{2}$ loads along with their corresponding fitted model. The loads are indexed by their equilibrium H$_{2}$ pressures. The upper panel corresponds to loads at 50 K and the lower at 77 K.}
\end{figure}
\noindent

\subsection{Loading the sample with Hydrogen}
Once equipped with a faithful modeling of the pristine sample, we could address the modeling of the effect of H$_{2}$ loading. The results of such an exercise are summarized in Fig.~\ref{NDH2loads}. As expected, the quality of the fits are not so good as in the pristine case (particularly in the shoulder range) but good enough to capture well the peak structure. A subtle shift of the peaks towards lower $Q$'s is observed, indicative of an expansion. At the same time, the strong inelastic effect makes the peak less and less pronounced with increasing load, overcoming the contrast enhancement expected from the negative value of the hydrogen neutron scattering length. At the maximum load, i.e., 4.2 MPa at 50 K the peak is almost a shoulder. Such a strong inelasticity effect hinders the ascertainment of the bundle expansion by a direct assignment of the peak location. Notice also that at 50 K the high of the peak eventually decreases with increasing load. This is evidence of partial conversion from ortho to para H$_{2}$ as a consequence of the interaction with carbon atoms. Essentially it amounts to a reducing overall factor not affecting peak shape.

\begin{figure}[t]
\centering
\includegraphics[width=0.48 \textwidth]{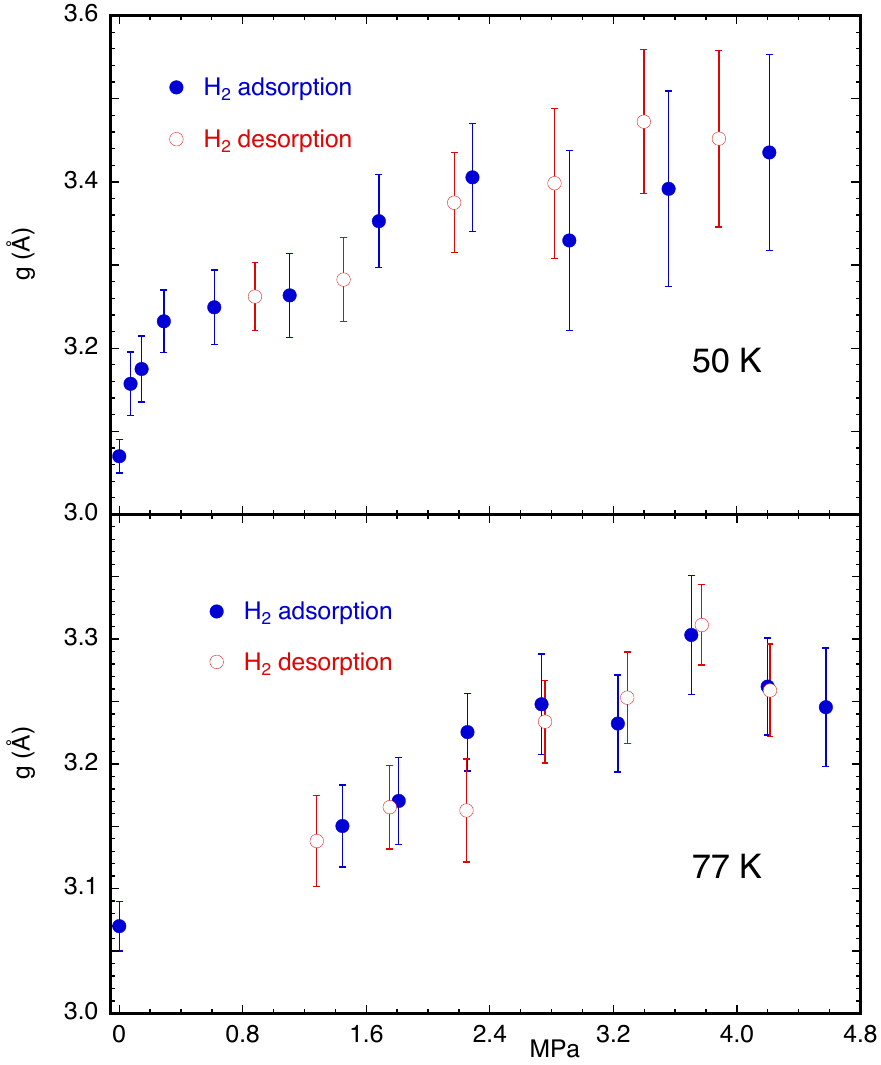}
\caption{\label{gaps} The evolution of the gap between nanotubes in a H$_{2}$ adsorption/desorption loop. The upper panel corresponds to loads at 50 K and the lower at 77 K.}
\end{figure}

The corresponding evolution with the H$_{2}$ load of the gap between the DWCNts is represented in Fig.\ref{gaps} for the complete loops of increasing/decreasing load at both 77 K and the posterior measurements at 50 K. Indeed, the results shown are a direct confirmation of the expansion of the DWCNT bundles. At 77 K the expansion shows initially a linear trend but eventually saturates to $\sim$  3.1 Å at $\sim$ 3 MPa. The same saturation is observed at 50 K, this time up to $\sim$ 3.4 Å, again at $\sim$ 3 MPa. Saturation in the gap growing is indicative of completion of the adsorption in all the accesible bundle interstitials (at around 3 MPa). Notice that the initial adsorption at 50 K follows a much more stepped trend than that at 77 K. Such behavior is not so surprising since loads are significantly larger owing to the lower temperature (see \cite{Cabrillo2024}). However, the final desorption of the H$_{2}$ at 77 K, due to the temporal restrictions of neutron scattering experiments, was made just by vacuum pumping without He purging. Such a purging is recommended in microporous samples. As a consequence probably there were some initial residual H$_{2}$ load in the 50 K measurements inducing a small vertical offset that exaggerates the slope of the initial load trend. Notice, also that because of our large error bars we cannot ascertain a possible structural hysteresis associated to that in the H$_{2}$ adsorption. As a matter of fact, the simpler approach to study the expansion by determining the location of the peak maximum, $Q_m$, as function of pressure, yields linear trends all the way up to the maximum pressures at both temperatures, that is, $Q_m$ is dominated by the deformation of the peak induced by the strong inelastic background.

Apart from confirming unequivocally that the hydrogen enters the interstitials, we see how the structure is flexible enough to expand up to $\sim$ 7 \% at 77 K, and up to $\sim$ 10 \% at 50K. The question arises of the role, if any, of the bundle expansion in the development of the adsorption hysteresis. 

\subsection{Atomistic numerical simulations}
In order to gain physical insight we first tried the most realistic approach, i.e., molecular dynamics (MD) simulations of a DWCNT bundle as described in section \ref{CSa}. As mentioned above, we estimated the displacement of the carbon atoms from this simulations. However, our main aim here was to follow the process of adsorption at a molecular level. Unfortunately, during the whole battery of simulations with pressures ranging up to 10 MPa, temperatures up to 800 K and times up to 150 ns, we could not observe any H$_{2}$ molecule get into the bundle in any of the MD simulations. 

At this junction, we reduced drastically the computational load by switching to a two dimensional (2D) analogue. The rational behind is to profit of the translational symmetry of the bundle along any of the axes of the nanotubes forming it. The movement of a molecule within an interstitial along any such axis is essentially brownian which tell us that any "trapping" of the molecules that could give raise to metastable adsorption is related with the transitions among interstitials. In a cross-section of the bundle the structure of the channels between interstitials is preserved. Besides, in the real system the space available in an interstitial is restricted to chunks limited by defects, torsional/curvature effects, amorphous carbon debris, etc.. The molecules eventually will fill the chunks so that at high loads the restrictions to the molecular movements do not depart so much from those in the 2D topology of a bundle cross-section. Altogether makes a 2D analogue made out of disks reproducing the cross-section of a bundle a promising model to study, albeit qualitatively, the adsorption hysteresis in the real system. 

\subsection{Microscopic origin of the adsorption hysteresis}
\begin{figure}
\centering
\includegraphics[width=0.48 \textwidth]{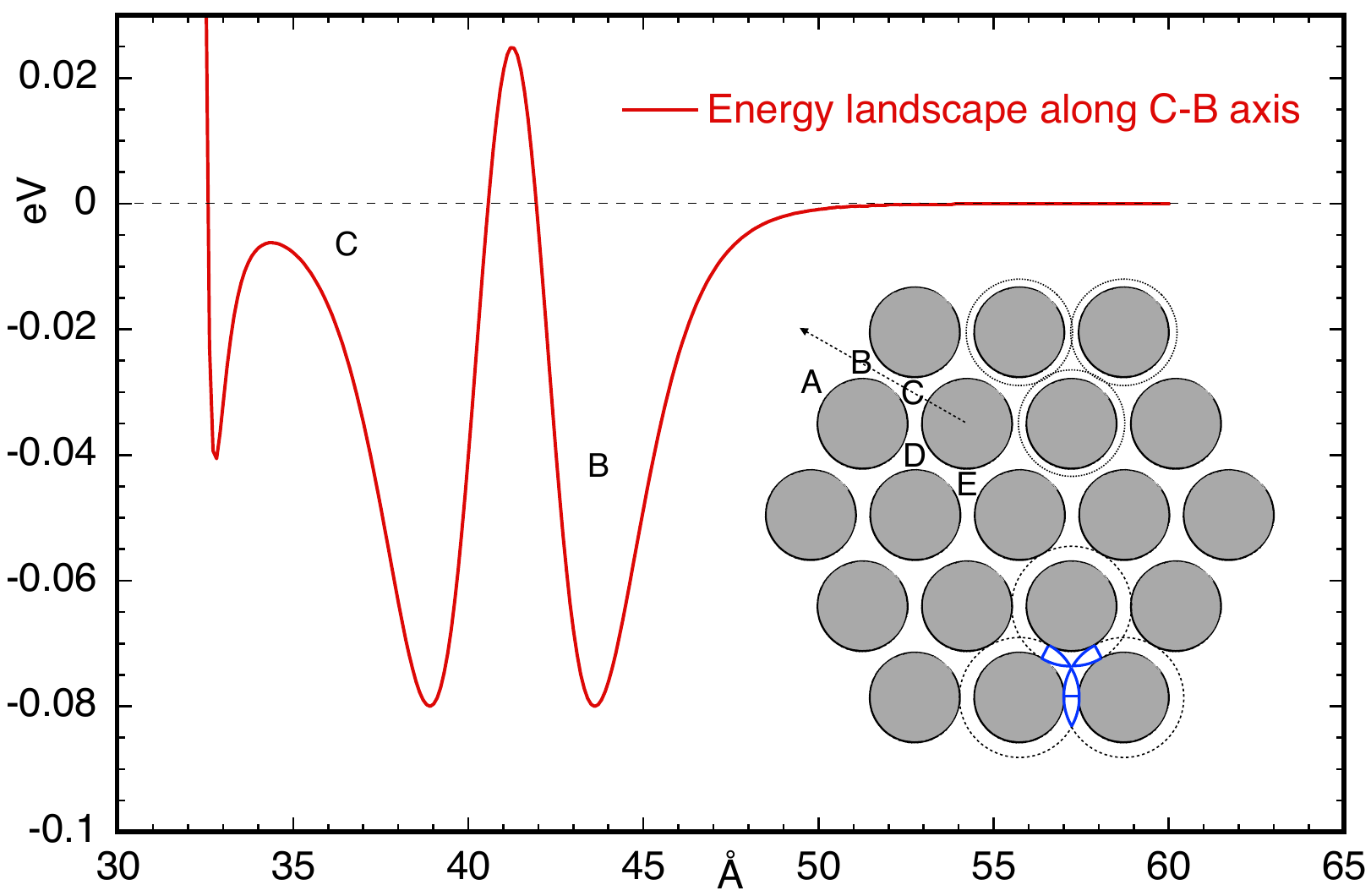}
\caption{\label{2DScheme} Potential energy in the 2D analogue bundle (shown in the inset) along the path marked with the dotted arrow. The  bundle corresponds to the rigid configuration with $\sigma = 1.72$ Å. Inset: the various qualitatively different adsorption regions are labeled with capital letters. The dotted circles around three of the bundle upper disks corresponds to the minimum of the H$_{2}$/nanotube interaction. The blue curved triangles in three of the bundle lower disks illustrates the areas defined as the loci of the minima at the sides of the energy barriers between interstitials (see main text).}
\end{figure}

The inset of figure \ref{2DScheme} displays the 2D analogue of a DWCNT bundle, i.e., a bundle of disks as described in \ref{CS2D}. The size chosen is the minimum that present all the adsorption regions that can be considered qualitatively different, regions labeled with capital letters. Its size is also very near to the optimum size obtained in the fits of the pristine sample (see inset in Fig~\ref{NDemptyModel}). 

The meagre computational load of the 2D analogue has allowed us to simulate adsorption measurements through MD, suitable to study metastable behavior since MD follows a realistic microscopic dynamics. Obviously, after times long enough fluctuations will provoke any metastable adsorption to decay towards the thermodynamic equilibrium so that strictly speaking stability is defined up to an arbitrary time ($\sim$ 10 ns) during which the gas density in the outer part of the simulation box do not change appreciably. When needed, the fundamental thermodynamic sates has been reached through a Monte Carlo technique allowing large molecular jumps (see section \ref{CS2D}).  

\begin{figure*}
\centering
\includegraphics[width=1.00 \textwidth]{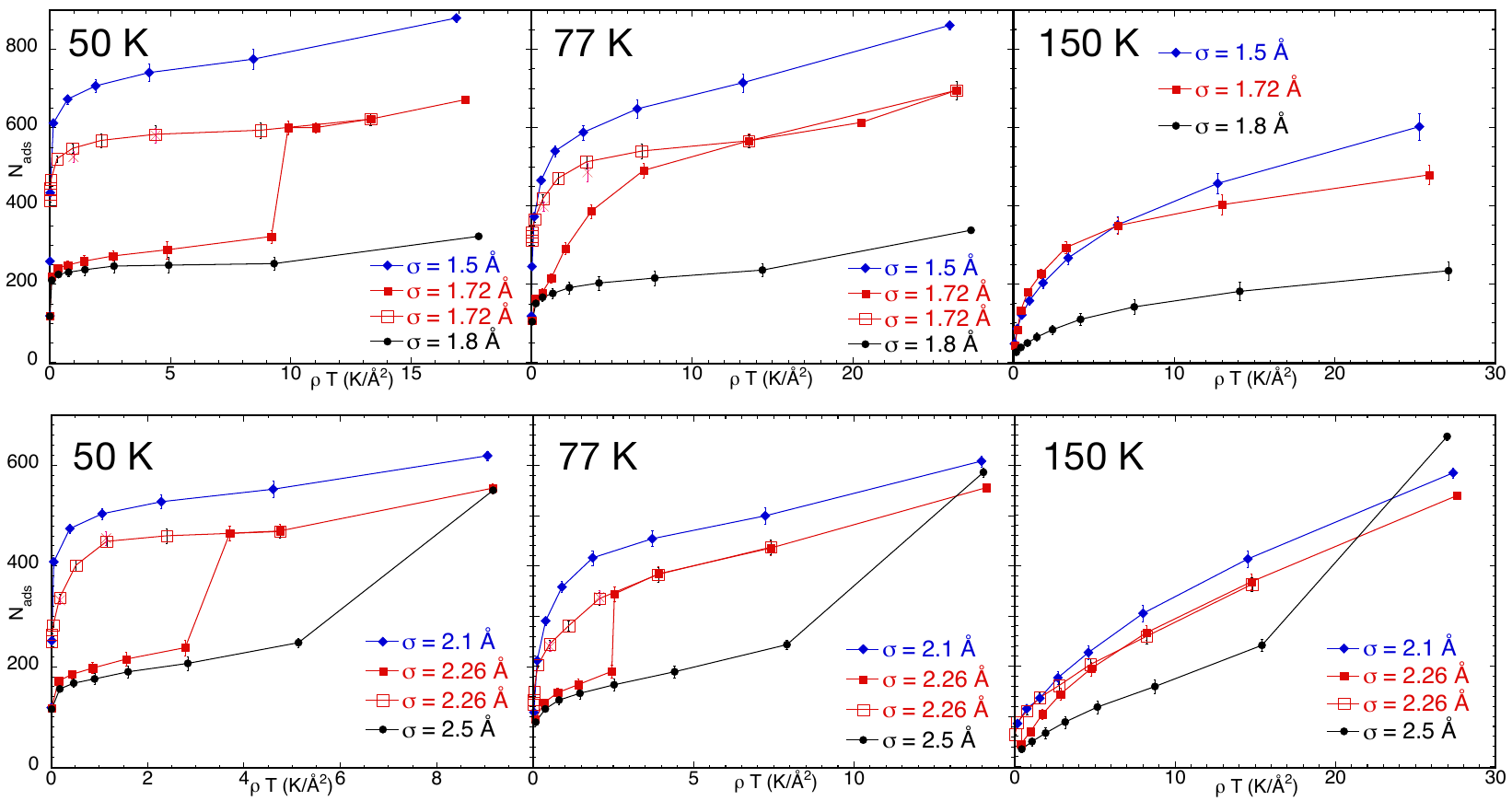}
\caption{\label{2DIsotherms} 2D adsorption isotherms at 50 K, 77 K and 150 K as a function of pressure (K/Å$^{2}$ per molecule). Upper panels correspond to fixed bundles (rigid disks fixed in the plane). Lower panels corresponds to flexible bundles (rigid disks bonded by harmonic oscillators). Solid symbols mark adsorption while hollow ones indicate desorption. Crosses within the hysteresis loops at 50 K and 77 K corresponds to the true thermodynamic equilibrium adsorptions.}
\end{figure*}

As described in the introduction, adsorption hysteresis is ascribed to deformable structures that reacts to the molecular insertion and/or to pore windows of molecular size. Consequently, as detailed in \ref{CS2D}, we have made two types of simulations, one with bundles made out of disks fixed in the plane and one where the disks are bonded with its first neighbors by harmonic oscillator interactions. Following our previous experiment, the studied temperatures are 50 K, 77 K, and 150 K. The key parameter is the core radius of the ``molecule" as expressed by the $\sigma$ parameter of the corresponding Lennar-Jones interaction between both, the ``molecular" disks and the ``molecules" with the ``bundle" disks. The isothermal adsorption results are summarized in Fig~\ref{2DIsotherms}. Hysteresis loops have been found in all cases but the fixed bundle at 150 K. 
\noindent

Centering first in the fixed bundle cases (upper panels in Fig.~\ref{2DIsotherms}), indeed we got hysteresis for very specific molecular radius at 50 K and 77 K. As shown in the figure at $\sigma = 1.72$ Å (to be compared with half the gap between the bundle disks, i.e., 1.7 Å) we got clear hysteresis loops. The specificity is high with the loop disappearing within $\pm 0.01$ Å around 1.72. That is, metastable adsorption happens to a size slightly greater than that of the porous so that a barrier should develop at the bundle entrance. Indeed, the calculated energy along an entrance channel for $\sigma = 1.72$ Å shows such a barrier as illustrated in Fig.~\ref{2DScheme}. The cases $\sigma = 1.5$ Å and $\sigma = 1.8$ Å serve as references where the molecule is to small (it goes in and out the bundle easily) or to big (do not enter the bundle at all) to observe hysteresis. The desorption points superimpose onto the adsorption ones in this reversible cases and are not plotted explicitly. At 150 K, however, we could not find any metastable loop. As expected hysteresis loops are also observed in the case of flexible bundles but now at a remarkably larger molecular sizes. That is, thermal vibrations amounts to larger effective porous sizes. Besides, a small but discernible loop is now observed at 150 K qualitatively reproducing the experimental results. It is also observed an anomalously high adsorption at the highest pressures and largest molecular sizes. It is an artifact owing to the deformable bundle structure. At the highest pressures (corresponding to crystalline molecular arrangements) the molecules still do not enter into the bundle but compress it appreciable so that the death area is reduced with respect to that used to compute the adsorption (taken at zero pressure). This not an artifact of our simulations but happens equally in real  experiments once the adsorbate structure can deform changing its death volume. However, at least for this particular structure the effect at the highest densities its greatly exaggerated since the relevant pressures are unrealistically high for an experiment. We have also chosen some specific points within the hysteresis loops at 50 K and 77 K and calculated the true  thermodynamic equilibrium adsorption by means of a large molecular jumps Monte Carlo algorithm (see section \ref{CS2D}). This equilibrium points (see red crosses Fig.~\ref{2DIsotherms}) reveals a desorption at equilibrium in the flexible bundles and only marginally out of equilibrium in the rigid case. That is, the deviation within the loops respect to the true thermodynamic equilibrium happens in its vast majority in the adsorption branch. We do not enter here into the thermodynamic nature of the metastable branches, i.e., whether they convey a real minimum of the free energy or arise as a consequence of purely kinetic effects.

\begin{figure}
\centering
\includegraphics[width=0.48 \textwidth]{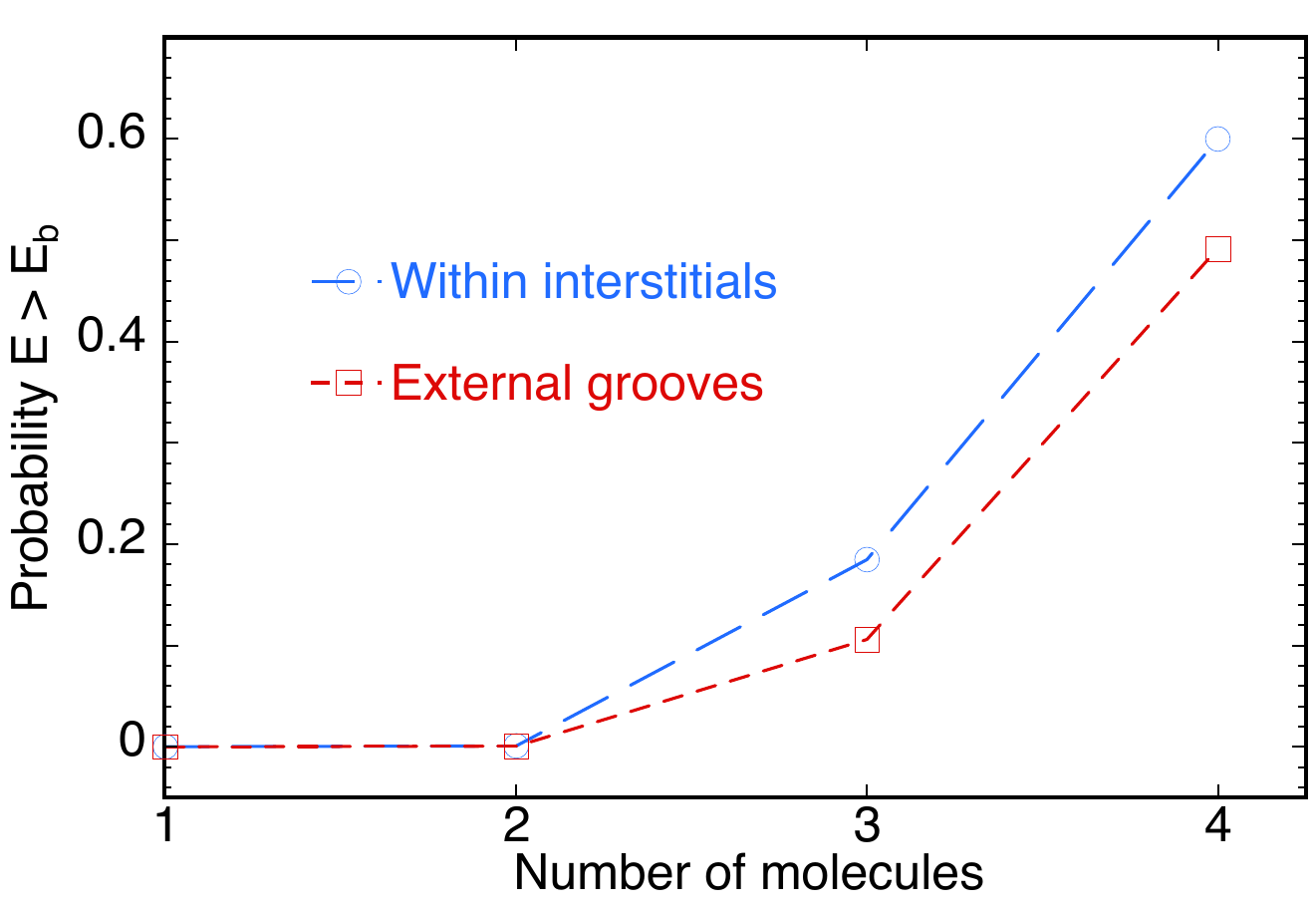}
\caption{\label{Pstate*} The probability of at least one molecule located around an energy minimum (blue curved triangles in Fig.\ref{2DScheme}) having an energy above the barrier between nanotubes as a function of the local load.}
\end{figure}

After all of these observations the question about the microscopic origin of the hysteresis arises naturally. Let us first center at the rigid bundle case. In Fig.~\ref{2DScheme} the energy along an entrance bundle path is displayed. We see that the $\sigma$ value for hysteresis (1.72 Å) is such that the loci of minimum interaction energy (dotted circles in Fig.~\ref{2DScheme}) slightly intercept each other at the middle of the channel between interstitials giving rise to an energy barrier ($\Delta E \simeq 0.1$ eV $\sim$ 1200 K) separating two equally depth potential wells equally contributed by the interaction of the two adjacent disks. This scheme of a symmetric potential doble well repeats in all the internal channels between the interstitials the difference among the various adsorption regions being only in the topology of neighboring interstitials. Such energy barrier structure suffices to understand how metastable adsorption emerges in this simple system. At low pressures, the molecules first accumulate at the strongly attractive B regions. At the studied temperatures they have not enough energy to surpass the energy barrier between B and C regions. But as the load increases and the B regions become populated, the energy of a molecule located there also increases, surpassing the energy barrier, and eventually entering regions C. To put the matter in more quantitative terms we have calculated the probability of a molecule being in an activated state, i.e., with an energy above the energy barrier (0.025 eV), as a function of the local load at a region nearby an energy minima. In Fig.~\ref{2DScheme} the chosen regions are depicted as blue curved triangles, one in each B region and three in each C, D and E regions. The probability of at least one molecule within a given region of being in an activated state as a function of the number of molecules in such a region are depicted in Fig.\ref{Pstate*}. There, the probabilities corresponding to all the external grooves (B regions) and the interstitials (regions C, D and E) has been aggregated. In both cases, a significant probability switch on at a local load of three molecules. The same process will repeat hierarchically filling regions D and E. In effect we have a gate/closing opening process dependent on the local load of molecules. This loading-dependent diffusion in nanoporous systems has been previously studied under the perspective of (corrected) transition state theory \cite{Beerdsen2004,Dubbeldam2005} and explicitly proposed in the context of hysteretic adsorption hysteresis \cite{Sircar2011}. The general idea of restricted diffusion in small-pore system has been also associated with hysteretic adsorption in rigid nanostructures \cite{Kang2010}.

We see then, that the bundle expansion under the hydrogen adsorption is not essential for the hysteretic behavior. However, the fact that at 150 K the hysteresis shows up only in the flexible case strongly suggests that the adsorption induced deformation of the nanostructure must have a significant effect into the hysteretic behavior. Indeed, a microscopic mechanism capable of inducing hysteresis is quite clear. Larger gaps imply a lower energy barrier between interstitials and therefore Fig.~\ref{gaps} describe a H$_{2}$ loading-dependent gate opening/closing mechanism. Such an effect becomes manifest in the simulations at high enough loads as shown in the supplementary material video \cite{SimulationMovie}. There, simulations at low and high loads are compared. It is seen how the gap size is larger at the high load so that the molecular movements between interstitials is somehow ``fluidized", transitions between interstitials being observed in short time scales. This mechanism is of a different nature than that of the previous one in the sense that in the depends in the cooperative interaction of the molecules onto the nanostructure (whose reaction in turn, depends on the internal nanostructure interactions) while the previous depends in the cooperative interactions of the molecules between themselves. Nevertheless, both depend in the formation of energy barriers that come from the interaction between the molecules and the nanostructure and in this sense they both are related. Notice, that the tuning of the involved interactions to give rise to the metastability is tight, as revealed by the narrow range in which hysteresis happens in the rigid structure simulations. Under this perspective, it is not surprising the lack of observations of the phenomena in other kinds of nanotube bundles.

The found microscopic mechanisms do not distinguish between adsorption and desorption. The observed difference in branch metastability where the desorption branches are at best marginally metastable comes from the asymmetry in the initial conditions. To make explicit as much as possible the hysteresis loops we have reached very high loads where the true thermodynamic equilibrium is reached. Desorption starts at equilibrium and at loads facilitating the diffusion of the molecules through the interstitials while adsorption begins with conditions hampering it. Notice also that the smallness of our system where a high interstitial load amounts to just $\sim$ 7 molecules gives a small space to play with more realistic conditions. 
\noindent

A last experimental observation was the visual inspection of the sample after being submitted to the H$_{2}$ loads at 77 K and 50 K. The pristine sample was a rather thick, somewhat fluffy, black fabric. However, after the diffraction experiments, where the sample was submitted up to around 5 MPa, the aspect of the sample changed to a thin compact black fabric. In fact, posterior adsorption measurements resulted in small results of no practical interest. The process made the bundles to coalesce eventually drastically reducing the available surface for adsorption. Practical values of H$_{2}$ physisorption at supercritical temperatures implies stringent requirements. Here, not only the microporosity but the meso and macro counterparts must combine adequately. Besides, the needed nano-structuration is in many cases too delicate to bear the rather harsh conditions of the practical implementations, as happens here. So, from the practical point of view the studied material would require rather stringent enhancements, namely, nanotube diameter optimization, bundle size optimization, some kind of functionalization to avoid bundle coalescence, etc. At any rate, we have here a topology of the nanostructured material leading to hysteresis that departs from those typical of MOF systems where in many cases the topology of barriers is linear.

 \section{Summary and conclusions}\label{SandC}
We have centered our efforts in elucidating the microscopic origin of the hysteresis recently discovered in the hydrogen adsorption by a nanostructured material made out of DWCNT bundles. Firstly, we have quantitatively characterized the structural evolution of the sample owing to hydrogen adsorption by means of neutron diffraction and posterior detailed modeling of the signal. An expansion of the bundle with increasing H$_{2}$ load has been revealed, blatantly showing the interstitials among nanotubes forming a bundle as the main adsorption regions. Secondly, we have taken advantage of the translational symmetry of the nanotube bundles along the nanotube axes to develop a two dimensional analogue of the system that allowed us numerical simulations of the adsorption processes in reasonable computer times.

Two microscopic mechanisms responsible for the hysteretic behavior emerged out of these efforts both based in the formation of energy barriers of the adequate high in the channels between the interstitials among the nanotubes. On one side, owing to the interaction among the molecules themselves, those energy barriers become surmountable as the H$_{2}$ load increases easing the molecular transitions between interstitials. It has a static character in the sense that no dynamics of the nanostructure is needed. On the other, the observed expansion of the bundles owing to the cooperative interaction of the hydrogen molecules onto the nanotube bundles lead to a decrease of the energy barriers again easing the molecular transitions between interstitials. In essence, two H$_{2}$-load dependent gate opening/closing mechanisms lead to the adsorption hysteresis.

\section*{Acknowledgements}
Financial support for this work has been secured through Grant  TED2021-129457B-I00 funded by MCIN/AEI/10.13039/501100011033 and the European Union NextGenerationEU/PRTR and by Ministerio de Ciencia, Innovaci\'on y Universidades (MCIU), under Grant No. PID2023-146442NB-I00 (AEI/FEDER, EU). We thank the Pressure and Furnace Team of the ISIS Neutron and Muon Source (Didcot, UK) for the excellent support during the preparation and realization of the neutron scattering experiments. We also acknowledge the continued support received from the IKUR Strategy under the collaboration agreement between Ikerbasque Foundation and the Materials Physics Center on behalf of the Department of Education of the Basque Government.






\bibliographystyle{model3-num-names}

\bibliography{/Users/carlos/Dropbox/Trabajo/Documentos/ArtiExt/BibTex/cc}

\end{document}